\documentclass[12pt,preprint]{aastex}

\IfFileExists{srcltx.sty}{\usepackage[active]{srcltx}}

\slugcomment{Submitted June 11, 2010; revised September 2, 2010}

\shorttitle{GAMMA-RAY HALOS AROUND AGN}
\shortauthors{ANDO \& KUSENKO}

\begin{document}

\title{Evidence for Gamma-Ray Halos Around Active Galactic Nuclei and
the First Measurement of Intergalactic Magnetic Fields}

\author{Shin'ichiro Ando}
\affil{California Institute of Technology, Mail Code 350-17, Pasadena,
 CA 91125, USA}
\email{ando@caltech.edu}
\and
\author{Alexander Kusenko}
\affil{Department of Physics \& Astronomy, University of California, Los
 Angeles, CA 90095, USA}
\affil{Institute for the Physics \& Mathematics of the Universe,
 University of Tokyo, Kashiwa, Chiba 277-8568, Japan}
\email{kusenko@ucla.edu}

\begin{abstract}
Intergalactic magnetic fields (IGMF) can cause the appearance of halos
 around the gamma-ray images of distant objects because an
 electromagnetic cascade initiated by a high-energy gamma-ray
 interaction with the photon background is broadened by magnetic
 deflections.
We report evidence of such gamma-ray halos in the stacked
 images of the 170 brightest active galactic nuclei (AGN) in the
 11-month source catalog of the \textit{Fermi} Gamma-Ray Space
 Telescope.
Excess over point spread function in the surface brightness profile is
 statistically significant at $3.5\sigma$ (99.95\% confidence level),
 for the nearby, hard population of AGN.
The halo size and brightness are consistent with IGMF, $B_{\rm
 IGMF}\approx 10^{-15} ~ {\rm G}$.
The knowledge of IGMF will facilitate the future gamma-ray and
 charged-particle astronomy. 
Furthermore, since IGMF are likely to originate from the primordial seed
 fields created shortly after the Big Bang, this potentially opens a new
 window on the origin of cosmological magnetic fields, inflation, and
 the phase transitions in the early Universe.
\end{abstract}


\section{Introduction}

%%% Introduction of IGMF (not including gamma-ray stuff) %%%
Intergalactic magnetic fields (IGMF) had not been measured until now,
despite their importance for gamma-ray and cosmic-ray astronomy and
their likely connection to the primordial fields that could have seeded
the stronger magnetic fields observed in galaxies, Sun, and Earth.
This is because IGMF are too small for conventional astronomical probes,
such as Zeeman splitting or Faraday rotation.
Unlike the fields in galaxies, which are believed to have been amplified
by the dynamo action of the large-scale convective motions
of gas, the fields in voids remain low, close to their primordial
values modified only by the relatively small contribution of the fields 
leaking out of galaxies
\citep{Kronberg1994,Grasso2001,Widrow2002,Kulsrud2007}.
The observational and theoretical upper bounds on IGMF constrain their
magnitudes to be below $10^{-9}$~G \citep*{1997PhRvL..78.3610B}, whereas
any value above $\sim$10$^{-30}$~G is sufficient to explain the $\sim
\mu$G Galactic magnetic fields generation by the dynamo mechanism
\citep*{1999PhRvD..60b1301D}.

%%% Introduction of secondary photon mechanism %%%
One can detect such extremely weak fields using high-energy gamma
rays \citep*{Aharonian1994,Plaga1995}.
Very energetic photons emitted from active galactic nuclei (AGN) or
other strong sources produce pairs of electrons and positrons in their
interactions with the extragalactic background light (EBL).
These pairs up-scatter the cosmic microwave background photons
to high energies, giving rise to an electromagnetic cascade, and the
photons from the cascade are detected by gamma-ray telescopes, such as
{\it Fermi}.
Since the trajectories of electrons and positrons in the cascade are
affected by magnetic fields, a gamma-ray image of AGN is expected to
exhibit a halo of secondary photons around a bright central
point-like source \citep{Aharonian1994,Dolag2009,Neronov2009}.
The central image is expected to be composed of photons emitted directly
from the source with energies below the pair production threshold.
In addition, delays in arrival times of the secondary photons can be
used to probe IGMF \citep{Plaga1995,Ando2004,Murase2008}.
Finally, at TeV energies, the secondary photons produced in interactions
of cosmic rays with EBL may have already been observed by the air
Cherenkov telescopes \citep{Essey2010a,Essey2010b}.

%%% Existing constraints by gamma-ray telescopes %%%
Thus far, in TeV range, HEGRA \citep{HEGRA} and MAGIC \citep{MAGIC}
did not detect any halo component of two bright blazars, Mrk~501 and Mrk~421,
and they set upper limits on the flux.
In particular, the analysis of MAGIC using gamma rays above 300~GeV
excludes some range of IGMF between $4\times 10^{-15}$ and $10^{-14}$~G.
Very recently, IGMF above $3\times 10^{-16}$~G were proposed as an
explanation of non-observation by {\it Fermi} of several AGN known to be
bright TeV sources (\citealt{Neronov2010}; see also
\citealt{2010MNRAS.tmpL..82T}).

In this {\it Letter}, we present evidence of extended images and of IGMF
at $3.5\sigma$ level, based on gamma-ray data collected by the Large Area
Telescope (LAT) onboard {\it Fermi}, in the energy range between 1~GeV
and 100~GeV.
It is consistent with pair-halo scenario with IGMF, $B_{\rm IGMF}
\approx 10^{-15} ~ {\rm G}$.
The knowledge on IGMF will facilitate the future gamma-ray and cosmic
ray astronomy, and it will open a new window on the origin of
cosmological magnetic fields \citep{Cornwall:1997ms}, inflation
\citep{Turner:1987bw,DiazGil:2007dy},  and the phase transitions in the
early Universe
\citep{Vachaspati:1991nm,Baym:1995fk,Vachaspati:2001nb}.

\section{Stacked Gamma-Ray Images of AGN}

%%% Introduction of analysis method %%%
The individual photon data as well as the 11-month source
catalog \citep{FermiSource} are now publicly
available.\footnote{\texttt{http://fermi.gsfc.nasa.gov/ssc/}}
Among $\sim$700 AGN in the {\it Fermi} AGN catalog \citep{FermiAGN}, we
select 170 AGN that are detected at more than $4.1\sigma$ in the highest
energy band, 10--100 GeV, and located at high Galactic latitudes, $|b| >
10\degr$.
These sources are likely to have a hard spectrum, and produce a large
number of TeV primary photons, which is necessary for the appearance of
the secondary halo.
Although each individual AGN produces too few photon counts, especially
in the highest-energy band, one can dramatically improve the statistics
by stacking all these 170 AGN maps.
We perform the analysis in three separate energy bands: 1--3 GeV, 3--10
GeV, and 10--100 GeV, which allows us to study the energy dependence of
the halos.
To obtain source and model maps, we use official {\em Science
Tools} made publicly available by the {\it Fermi} team.
The photons that we use in the AGN analysis are collected between
239557417~s and 268416079~s in the mission elapsed time (MET), and they
are of ``Diffuse'' class.

We use locations of AGN from the 11-month source catalog, i.e., those
obtained solely by the gamma-ray data.
This does not introduce any significant uncertainty of the stacked
images, because the localization accuracy using gamma rays is typically
much better than the size of PSF especially for hard AGN
\citep{FermiAGN}.

\begin{figure}
\begin{center}
\includegraphics[height=8cm]{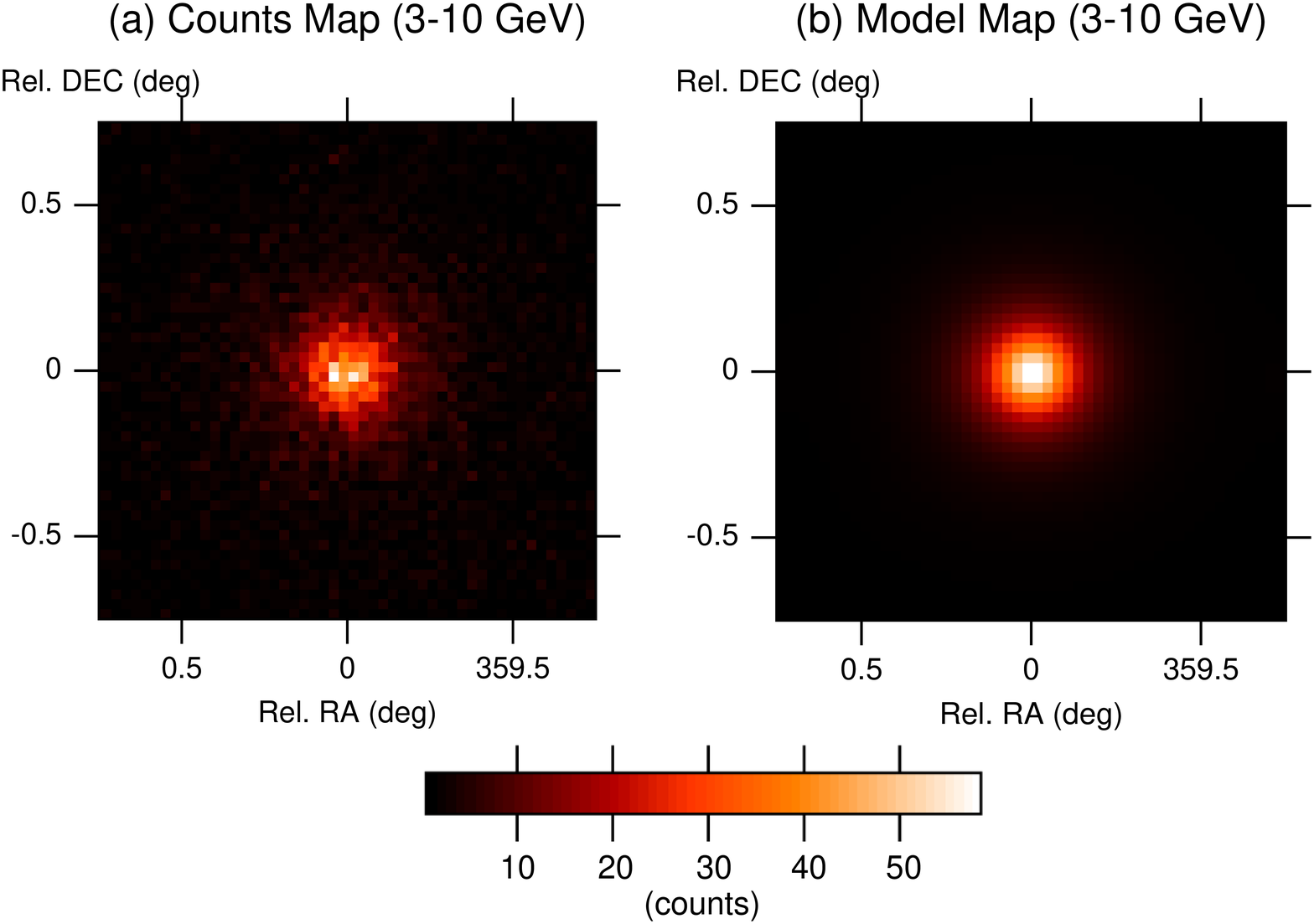}
\vspace{0.5cm}
\includegraphics[height=8cm]{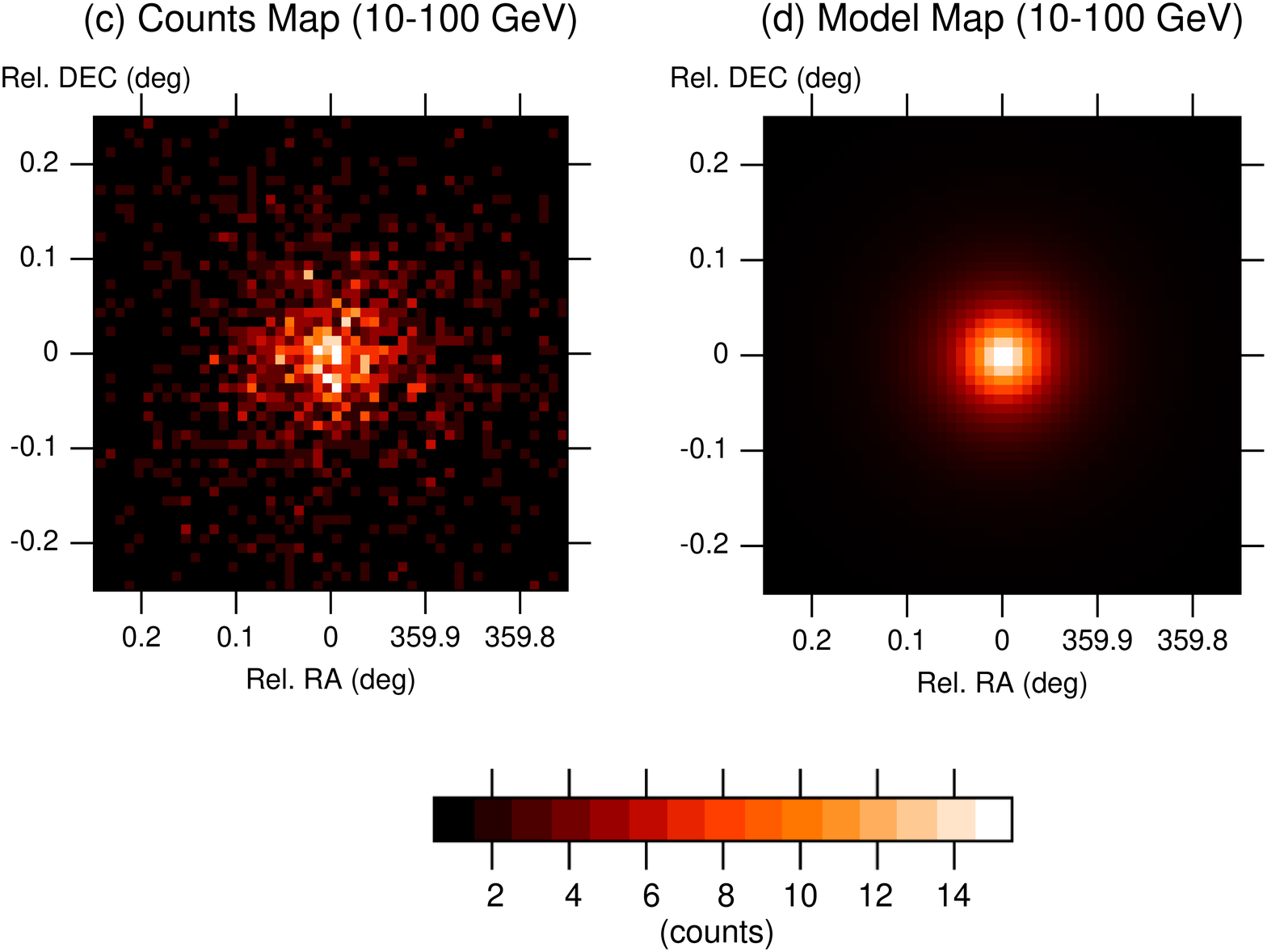}
\caption{Gamma-ray counts and point-source model maps of stacked
 170 bright AGN.  Upper and lower panels are for 3--10 GeV ({\it a} and {\it
 b}) and 10--100 GeV ({\it c} and {\it d}) bands, respectively.  Left
 panels ({\it a} and {\it c}) are the actual data counts of stacked
 170 AGN, and the right panels ({\it b} and {\it d}) show the
 ``best-fit'' point-source model (including backgrounds).  Pixel size is
 0.03$\degr$ (0.01$\degr$) for the 3--10 (10--100) GeV band.}
\label{fig:maps}
\end{center}
\end{figure}

%%% Stacked image map %%%
Figure~\ref{fig:maps} shows the gamma-ray count maps of stacked 170 AGN
and the ``best-fit'' point-source model generated with the {\it Fermi
Science Tools} as well as point-source catalog, smeared only by point
spread function (PSF) of LAT (we use the latest ``Pass6 version 3''
instrument response function of LAT).
It is evident that the counts map and the model map are not consistent
with each other, especially in the 10--100~GeV range.
%(We do not show the 1--3 GeV maps here, because there is no apparent
%visual difference between the counts and model maps.)

\section{Flux and Angular Extent of Halo Component}

%%% Analysis of stacked map %%%
We have performed maximum likelihood analysis assuming that, in addition
to the central point sources and diffuse backgrounds
\citep*{Strong2004,GB,EGB}, there is a third component, namely, the halo
component, whose spatial extent is given by the Gaussian distribution:
\begin{equation}
P_{\rm halo}(\theta^2|\theta_{\rm halo}^2) = \frac{2}{\pi \theta_{\rm
 halo}^2}
 \exp\left({-\frac{\theta^4}{\pi \theta_{\rm halo}^4}}\right),
\end{equation}
where $\theta$ is the angle from the map center, and $\theta_{\rm
halo}^2$ is the mean of $\theta^2$ over this distribution function,
$\theta_{\rm halo}^2 \equiv \langle \theta^2 \rangle$.
We fit the histogram of photon counts as a function of $\theta^2$
read from the maps by minimizing
\begin{equation}
\chi^2 = \sum_i \frac{1}{N_i}
 \left[ N_{\rm psf} P_{\rm psf}(\theta_i^2) + N_{\rm halo} P_{\rm
  halo}(\theta_i^2 | \theta_{\rm halo}^2) + N_{{\rm bg},i}
  -  N_i\right]^2,
\end{equation}
where $N_{\rm psf}$, $N_{\rm halo}$, and $\theta_{\rm halo}$ are
treated as free parameters.
The index $i$ refers to the $i$-th bin, $N_i$ is the total number of
events in this bin, $P_{\rm psf}$ is the normalized PSF, and $N_{{\rm
bg},i}$ is the events due to diffuse backgrounds.
We fix the backgrounds to the values at $\theta^2 = 2.025$--2.25 deg$^2$
and 0.233--0.25 deg$^2$ for 3--10 GeV and 10--100 GeV, respectively, in
the simulated maps, assuming that they are homogeneous.
%(This is a very good approximation because we average the backgrounds
%over many AGN images.)
Thus, $N_{\rm psf}$ and $N_{\rm halo}$ are the total numbers of photons
in the map attributed to the point source and the halo, respectively,
and $\theta_{\rm halo}$ is the apparent angular extent of the halo
component.

\begin{figure}
\begin{center}
\includegraphics[width=8cm]{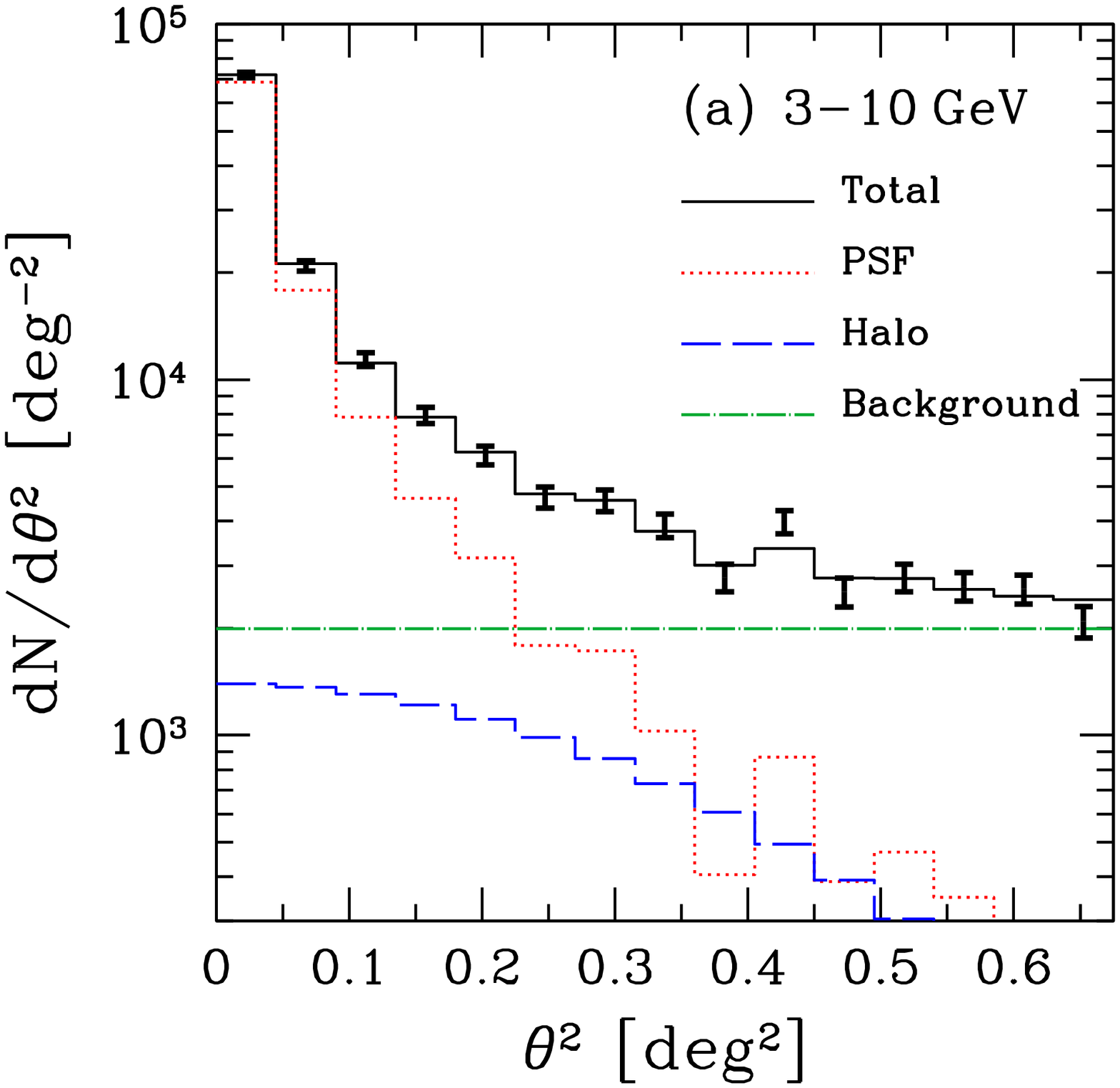}\\
\includegraphics[width=8cm]{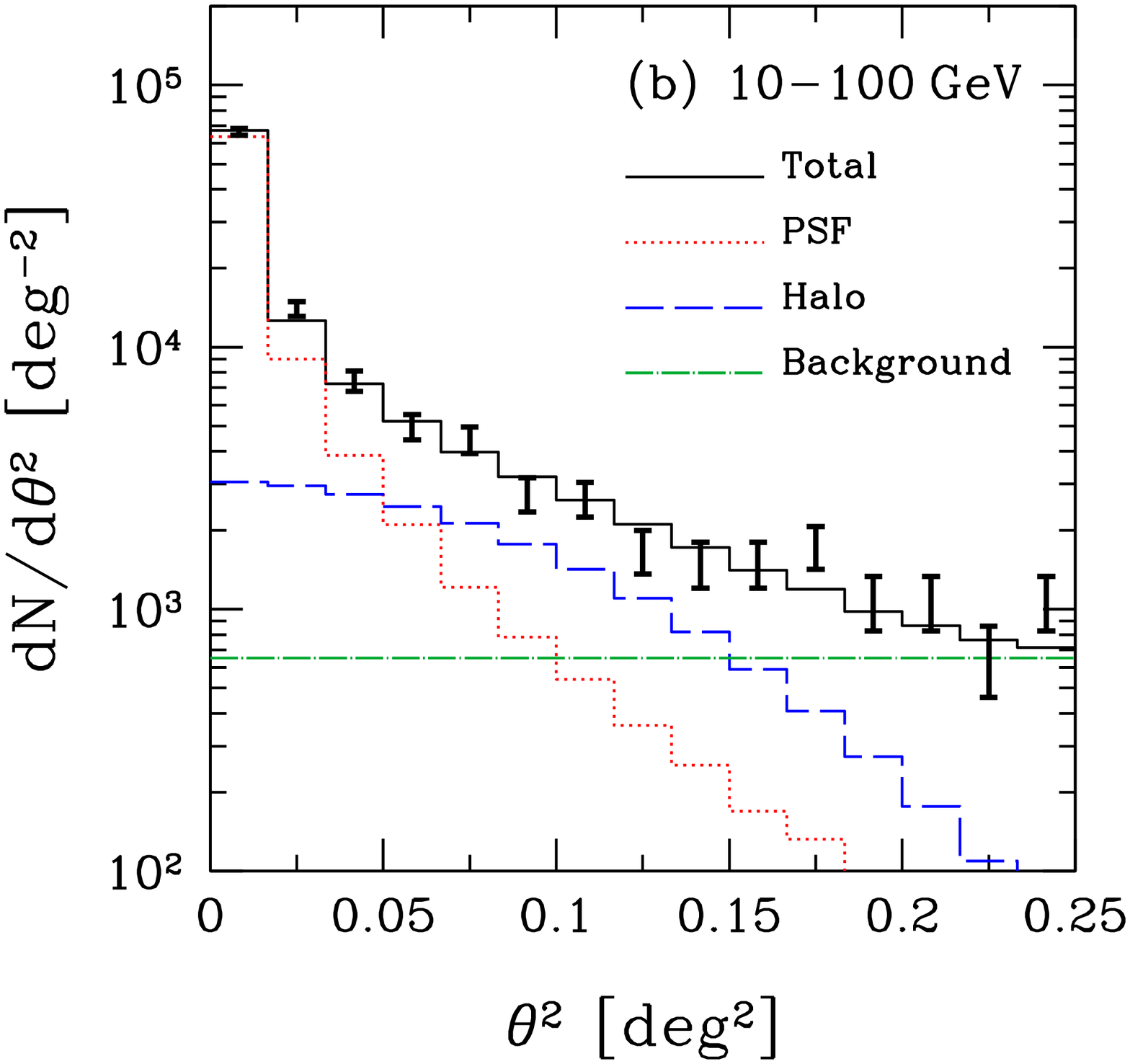}
\caption{Surface brightness profile of the stacked gamma-ray
 images.  Panels {\it a} and {\it b} are for 3--10 GeV and 10--100 GeV
 bands, respectively.  Points with error bars are the data, and solid
 histogram is the best-fit model.  The dotted, dashed, and dot-dashed
 histograms represent truly point-like source, halo component, and
 homogeneous diffuse background, respectively.}
\label{fig:bestfit}
\end{center}
\end{figure}

%%% Results of maximum likelihood analysis %%%
The inclusion of the halo component improves the fit significantly at
high energies.
The minimum $\chi^2$ over degree of freedom ($\nu$) is $\chi_{\rm
min}^2/\nu = 18.8 / 19$ and $13.3/12$ for 3--10 GeV and 10--100 GeV,
respectively.
In contrast, the ``best-fit'' point-source model, where $N_{\rm psf}$
and the background amplitude are treated as free parameters, gives
$\chi_{\rm min}^2 / \nu \simeq 66 / 20$ and $62 / 13$ for 3--10 GeV and
10--100 GeV, respectively.
This clearly shows that, even though we stack many AGN, this simple
Gaussian halo model gives a very good fit to the data.
The surface brightness profiles $dN/d\theta^2$ of the best-fit halo
model are juxtaposed with the data points in Fig.~\ref{fig:bestfit}.

%%% Allowed parameter regions %%%
In Fig.~\ref{fig:contour}{\it a}, we show the allowed regions of
$\theta_{\rm halo}$ and $f_{\rm halo}$ at 68\% and 95\% confidence
levels.
Here $f_{\rm halo}$ is the fraction of the halo photons, i.e., $f_{\rm
halo} \equiv N_{\rm halo} / (N_{\rm psf} + N_{\rm halo})$.
The best-fit values and $1\sigma$ statistical errors for these
parameters are $\theta_{\rm halo} = 0.49\pm 0.03\degr$ and $f_{\rm halo}
= 0.097 \pm 0.014$ for 3--10 GeV, and $\theta_{\rm halo} =
0.26 \pm 0.01\degr$ and $f_{\rm halo} = 0.20 \pm 0.02$ for
10--100 GeV.
For the lowest energy band, 1--3 GeV, only an upper limit on $f_{\rm
halo}$ is obtained, which is $f_{\rm halo} < 0.046$ at 95\% confidence
level.

\section{Eliminating Instrumental Effects}

\subsection{Dependence on redshifts and spectra}

\begin{figure}
\begin{center}
\includegraphics[width=8cm]{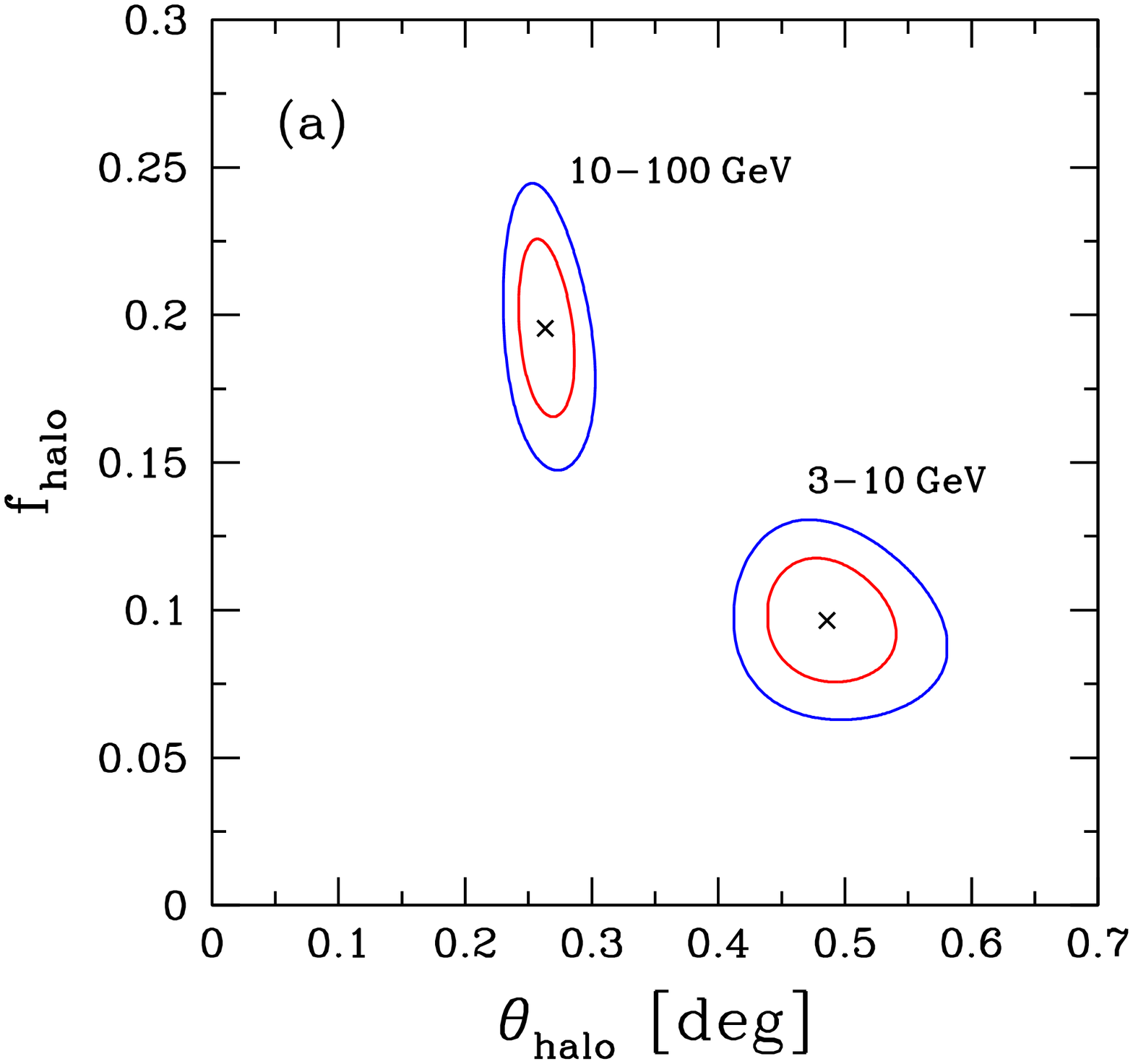}\\
\includegraphics[width=8cm]{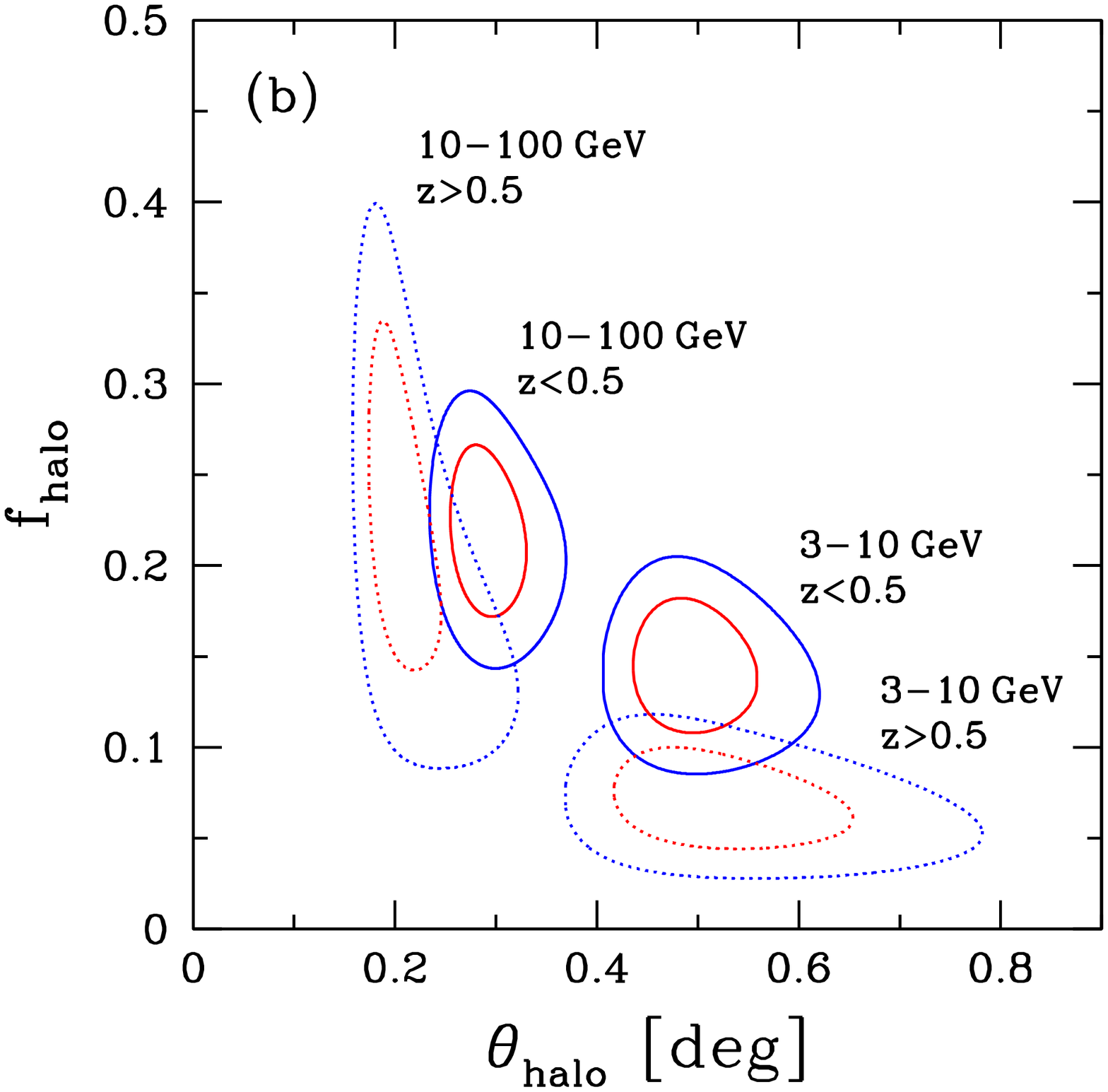}
\caption{Allowed regions of halo size $\theta_{\rm halo}$ and
 fraction of events from halo component $f_{\rm halo}$.  {\it a:}
 Contours are at 68\% and 95\% confidence level obtained with the sample
 of 170 AGN.  The best-fit values are marked by the crosses.
 Lower-right and upper-left contours are for 3--10 GeV and 10--100 GeV,
 respectively.  {\it b:} The same as {\it a} but for
 99 AGN with known redshifts (again, right and left contours are for
 3--10 GeV and 10--100 GeV, respectively). Allowed regions for nearby 57
 AGN ($z < 0.5$) are shown by solid contours, and those for distant 42
 AGN ($z > 0.5$) are by dotted contours.}
\label{fig:contour}
\end{center}
\end{figure}

%%% PSF discussion %%%
We discuss the possibility that these halos could be due to
some unknown instrumental effect, such as, for example, a possible
deviation of LAT PSF from its value measured in calibration prior to the
launch.
To exclude such a possibility, we first consider samples of AGN at
different redshifts.
%, because any instrumental effects should be redshift independent.
We divide the 99 AGN with known distances (out of total 170) into two
groups: a sample of 57 nearby AGN with $z < 0.5$, and a sample of 42
distant AGN with $0.5 < z < 2.5$, where $z$ is the redshift of the
source.
The allowed regions of $\theta_{\rm halo}$ and $f_{\rm halo}$ for these
two samples, both for 3--10 GeV and 10--100 GeV, are shown in
Fig.~\ref{fig:contour}{\it b}.
The statistically significant difference between the two populations
shown in this figure implies that at least some component of the halos
cannot be attributed to instrumental effects.

We also note that most AGN in the nearby sample at $z < 0.5$ (53 among
57) are classified as the hardest population of gamma-ray blazars, BL
Lac objects \citep{FermiAGN,Abdo2010}.
The distant sample, on the other hand, consists of 33 flat-spectrum
radio quasars (and 9 others), which are known to be softer
population.
The fact that we measured the brighter and more extended
additional components for the nearby/hard sample
(Fig.~\ref{fig:contour}{\it b}) is consistent with the pair-halo
scenario, because the harder AGN radiate more TeV photons that source
secondary halos as well as they are closer.
This cannot be easily understood as instrumental systematics, on the
other hand, because the true PSF size would have to be an increasing
function of energy, which is not the case; see discussions in the next
subsection.

As another independent test to rule out instrumental effects, we
considered a sample of 43 AGN from the same catalog \citep{FermiSource}, 
which produced no photons above 10~GeV but were detected in the
3--10~GeV band at more than $4.1\sigma$.
These sources are likely to have a softer spectrum, with a negligible
flux of primary photons above the pair production threshold.
In the absence of pair production, one expects to see no halos.
As expected, the best fit in the 3--10~GeV band is achieved for
$f_{\rm halo} = 0$, with an upper limit of $f_{\rm halo} < 0.1$ at 95\%
confidence level.

\subsection{Quantitative estimate of instrumental effects}

The two independent tests described above give one confidence that
instrumental effects cannot account for all the observed halos.
\citet{Neronov2010b} repeated our stacking analysis and found the
same anomalous excess in the 10--100 GeV band.  
However, they argue, ``most, if not all, of this excess is due to the
imperfect knowledge of the PSF for the back-converted gamma-rays.''
This argument is based on the observation that the extent of the Crab
pulsar is the same as that of AGN, and the excess is different between
front and back-converted photons.
While we agree with \citet{Neronov2010b} that it is good to perform
other independent tests, we shall show that their arguments fail 
to exclude the physical halos and overturn the statistical significance
of redshift and spectrum tests discussed above.
To this end, we have performed an alternative analysis, using the
observed Crab profile as a calibrated PSF template.
This confirms our initial conclusion and demonstrates that the halos are
indeed physical, at $3.5\sigma$ level.
For the analysis, we mainly focus on the 3--10 GeV band, because the
data have more statistical power than in 10--100 GeV, as well as the
pre-launch PSF is better calibrated at lower energies
\citep*{Burnett2009}.

%In this alternative analysis we mainly focus on the 3--10 GeV band,
%which has the following advantages over the 10--100 GeV band.
%First, the Crab pulsar can be regarded as a point source with a greater
%confidence.
%Second, the data have more statistical power, as there are about five
%times more photons.
%Finally, the pre-launch PSF is better calibrated at lower energies
%\citep*{Burnett2009}.

In Fig.~\ref{fig:crab}, we show surface brightness profiles of our
nearby and distant samples of AGN, where one can see clear difference
between the two populations of AGN.
In the same figure, we also plot the profile of Crab,\footnote{The Crab
profile is obtained from Diffuse-class data for ${\rm MET} =
239557417$--302034833~s.} which appears to be more consistent with the
distant AGN than the nearby set.
The backgrounds have been subtracted from the sources; they were
estimated based on the large angular regions, where the contributions
from both the point sources and halos are expected to be small.
We note that the excess of AGN over Crab seen in Fig.~4 was not found by
\citet{Neronov2010b}, who analyzed the data in the 10--100 GeV band,
which, as mentioned above, lacks statistical power in comparison with
the 3--10 GeV band used here.

To proceed with a quantitative analysis, we use this Crab profile as a
PSF model in this energy range, and regard Crab statistical errors as
systematic uncertainties of PSF.
For example, in angular bin $\theta^2 = 0.225$--$0.27$ deg$^2$, {\it
Fermi}-LAT received 22 photons from Crab, and the background is
estimated to be 1.9. 
This is interpreted as 24\% systematic uncertainty of PSF in this
particular bin.
This method is independent of our previous analysis and is free of any
uncertainties related to pre-launch PSF calibration.

A possible source of additional systematic uncertainties is an energy
dependence of PSF.
In general, gamma-ray spectra are different between AGN and pulsars, and
so are expected PSF sizes.
However, the detected spectra of both stacked AGN and Crab are well
approximated by a power law with similar indices; $dN/dE_\gamma
\propto E_\gamma^{-2.2}$ for nearby AGN ($z < 0.5$), and $\propto
E_\gamma^{-2.4}$ for Crab, where $E_\gamma$ is the gamma-ray energy.
To probe the spectrum dependence of PSF even further, we compare the
brightness profiles of {\it simulated} maps of nearby/hard and
distant/soft AGN.
%; even though this model turned out to have been
%mis-modeled, it will still give indication of magnitude of the effect.
Both profiles look very similar, while the profile of hard AGN is
slightly less extended.
In angular bin $\theta^2 = 0.225$--$0.27$ deg$^2$, these AGN profiles
differ only by 4.4\%, which is negligible compared with 24\% uncertainty
of the Crab-based PSF due to Crab statistics.
Finally, we note that the width of the PSF decreases with energy. 
We verified this by comparing Crab images in 3--5 GeV and 5--10 GeV
bands, and confirming that the former is broader than the latter.
Since the spectrum of the nearby AGN is harder than that of Crab, the
instrumental systematics can only make the AGN image sharper,
not broader, but the opposite is inferred from Fig.~\ref{fig:crab}.
Therefore, it is conservative to ignore the small systematic
uncertainties due to spectrum dependence of PSF.

Adopting the Crab profile as a calibrated PSF, we quantitatively
investigate the excess of nearby/hard AGN profiles identified at
$\theta^2 \gtrsim 0.2$~deg$^2$ in Fig.~\ref{fig:crab}.
The PSF and AGN profiles are normalized to each other such that they
give the same brightness in the innermost angular bin, $\theta^2 <
0.045$~deg$^2$.
The excess photon counts are $N_{\rm excess}^{(z<0.5)}(\theta^2 > 0.225
~\mathrm{deg^2}) = 125 \pm 30({\rm stat}) \pm 21({\rm sys})$.
By taking square root of quadratic sum of the statistical and systematic
errors as a total error, we find that this excess is of $3.5\sigma$
significance.
For the distant/soft AGN population, on the other hand, the excess is
$N_{\rm excess}^{(z>0.5)}(\theta^2 > 0.225 ~\mathrm{deg^2}) = -5 \pm
27({\rm stat}) \pm 29({\rm sys})$, consistent with null hypothesis.
Dividing $N_{\rm excess}$ by the total number of PSF counts, we obtain
the values of $f_{\rm halo}$ for both AGN populations:
\begin{equation}
f_{\rm halo} =
 \left\{
  \begin{array}{rr}
   0.073 \pm 0.017({\rm stat}) \pm 0.012({\rm sys}), & \mbox{for
    }z<0.5,\\
   -0.002 \pm 0.011({\rm stat}) \pm 0.012({\rm sys}), & \mbox{for
    }z>0.5.\\
  \end{array}
 \right.
\end{equation}
Clearly, this conclusion using the Crab-calibrated PSF agrees with that
based on the pre-launch calibration.

One can go even further and design two separate Crab-calibrated PSFs for
two classes of photons, namely those that convert in the front layer and
those in the back layer of the detector.
While all of these photons must be used in an analysis, allowing for the
differences in PSF offers yet another opportunity to find and eliminate
some unexpected instrumental effects.
To this end, we introduce another statistical quantity $\delta_{\rm
excess} \equiv N_{\rm excess}^{\rm front} / N_{\rm psf}^{\rm front} + 
N_{\rm excess}^{\rm back} / N_{\rm psf}^{\rm back}$, where all the $N$'s
with self-explanatory superscripts and subscripts refer to photon counts
at $\theta^2 > 0.225$ deg$^2$ after the homogeneous backgrounds were
subtracted.
This way, we explicitly include any PSF differences between front and
back-converted photons.  
The meaning of $\delta_{\rm excess}$ is clear: a value consistent with
zero corresponds to absence of physical halos.
We obtain $\delta_{\rm excess}^{(z<0.5)} = 1.4 \pm 0.5 ({\rm stat}) \pm
0.2 ({\rm sys})$, $2.7\sigma$ away from the null hypothesis.
We also find that the individual values of $\delta_{\rm excess}$ for the
front and back photons are consistent with each other, within errors.

We have also performed the same analysis for 10--100 GeV.
Here, we renormalized the Crab and AGN profiles using $\theta^2 < 0.025$
deg$^2$ bin, and counted the excess photons over Crab-calibrated PSF in
$\theta^2 = 0.075$--0.25 deg$^2$.
We obtain $N_{\rm excess}^{(z<0.5)} = 19 \pm 13 ({\rm stat}) \pm 15
({\rm sys})$ and $N_{\rm excess}^{(z>0.5)} = -3.6 \pm 9.1 ({\rm stat})
\pm 8.2 ({\rm sys})$.
The excess for nearby/hard AGN is found significant at $1\sigma$
level.

\section{Implications for Intergalactic Magnetic Fields}
\label{sec:IGMF}

%%% Interpretation %%%
We interpret the size of the halo $\theta_{\rm halo}$ of a few tenths of
degree, in terms of the secondary photon model, especially parameters
of IGMF.
A simple analytic model gives the following relation between these
quantities \citep{Neronov2009}:
\begin{equation}
\theta_{\rm halo} = 
 \left\{
  \begin{array}{ll}
   5\degr (1+z)^{-2} \tau^{-1}
    \left(\frac{E_\gamma}{10~{\rm GeV}}\right)^{-1}
    \left(\frac{B_{\rm IGMF}}{10^{-15} ~ {\rm G}}\right), & \mbox{for
    }\lambda_B \gg
    D_e, \\
   0.4\degr (1+z)^{-1/2} \tau^{-1}
    \left(\frac{E_\gamma}{10~{\rm GeV}}\right)^{-3/4}
    \left(\frac{B_{\rm IGMF}}{10^{-15} ~ {\rm G}}\right)
    \left(\frac{\lambda_B}{1 ~ {\rm kpc}}\right)^{1/2}, & \mbox{for
    }\lambda_B \ll
    D_e,
  \end{array}
 \right.
 \label{eq:theta_halo}
\end{equation}
where $\tau$ is the optical depth for the TeV photons that produce
halo gamma rays, $\lambda_B$ are the correlation length of IGMF, and
$D_e$ is the energy-loss length of the electrons and positrons produced
by primary TeV photons.
Because the lower-energy secondary photons originate from the less
energetic electrons and positrons that are deflected by larger angles
in IGMF, one expects a larger halo size $\theta_{\rm halo}$ at lower
energy.
As for the dependence on $\lambda_B$, if it is much longer than
$D_e$, then the charged particles can be regarded as propagating in
homogeneous magnetic fields (equivalent to infinite $\lambda_B$), and
therefore, the deflection angle is given by the ratio of $D_e$ and the
Larmor radius.
If $\lambda_B$ is much smaller than $D_e$, on the other hand, then the
electrons and positrons propagate by random walk, with deflections
proportional to $\lambda_B^{1/2}$.

\begin{figure}
\begin{center}
\includegraphics[width=8cm]{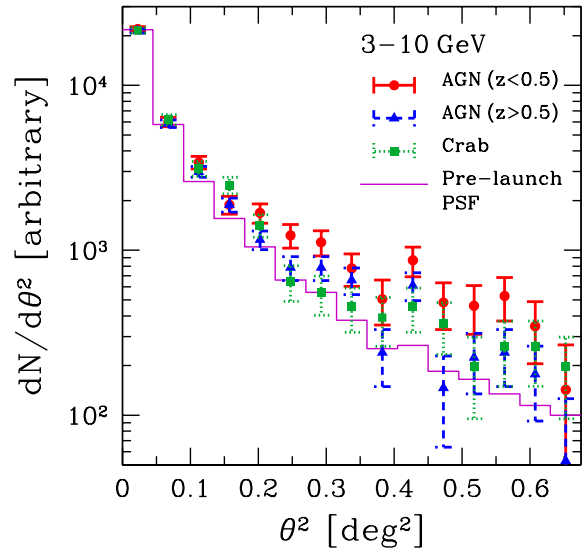}
\caption{Surface brightness profile of AGN and the Crab pulsar as well
 as pre-launch PSF (for Crab) in the 3--10 GeV band.  The diffuse
 backgrounds are subtracted.  The profiles are rescaled such that the
 innermost bin has the same brightness, and therefore, units in the
 vertical axis are arbitrary.}
\label{fig:crab}
\end{center}
\end{figure}

The halo size for the nearby AGN sample is $\theta_{\rm halo} \approx
0.5$--$0.8\degr$ (Fig.~\ref{fig:crab}).
Assuming $\tau \sim 1$--10 and using an average redshift of the nearby
AGN sample, $\langle z \rangle = 0.2$, the measured extent of the
halos is consistent with $B_{\rm IGMF} \approx 10^{-15} ~ {\rm G}$.
This is the first measurement of the strength of IGMF based on a
positive detection.
With the halo detection also in the 10--100 GeV band, we would be able
to constrain the correlation length by investigating the energy
dependence of $\theta_{\rm halo}$.

%%% No other realistic objects %%%
At the mean redshift of the nearby AGN sample, the observed halo size of
$\sim$0.5--0.8$\degr$ corresponds to 6--10 Mpc.
There are no known astrophysical sources capable of producing images of
such a large size.
Therefore, we conclude that the halos of the secondary photons provide
the only realistic explanation of the data.

\acknowledgments

S.A. was supported by Japan Society for Promotion of Science
 and partially by NASA through the Fermi GI Program grant NNX09AT74G. 
A.K. was supported by DOE grant DE-FG03-91ER40662 and NASA
 ATP grant NNX08AL48G.
A.K. thanks Aspen Center for Physics for hospitality.

\clearpage

%%
%% TABLES
%%
%% If there are any tables, put them here.
%%

\end{document}